\documentclass{aa}
\usepackage{xspace}
\usepackage{natbib}
\usepackage{graphicx}
\usepackage{amsmath}
\usepackage{amssymb}
\usepackage{url}
\usepackage{txfonts}
\usepackage{rotating}

\newcommand{\integral}{\textsl{INTEGRAL}\xspace}

\newcommand{\ginga}{\textsl{Ginga}\xspace}
\newcommand{\ca}{\mbox{$\sim$}}
\newcommand{\kevnxs}{\ensuremath{\text{ke\kern -0.09em V}}}
\newcommand{\mevnxs}{\ensuremath{\text{Me\kern -0.09em V}}}
\newcommand{\kev}{\kevnxs\xspace}

\newcommand{\err}[2]{\ensuremath{^{+#1}_{-#2}}\xspace}

\newcommand{\Msun}{\ensuremath{\mbox{M}_\odot}\xspace}

\newcommand{\xte}{\textsl{RXTE}\xspace}

\newcommand{\asm}{\textsl{ASM}\xspace}

\newcommand{\ev}{e\kern -0.11em V\xspace}

\newcommand{\vela}{Vela~X-1\xspace}
\newcommand{\ecut}{\ensuremath{E_{\text{Cut}}}\xspace}

\newcommand{\ecyc}{\ensuremath{E_{\text{C}}}\xspace}

\newcommand{\V}{V\,0332\ensuremath{+}53\xspace}
\newcommand{\grad}{\ensuremath{^\circ}\xspace}

\begin{document}
\title{\integral observation of \V in outburst}

\author{I. Kreykenbohm\inst{1,2} \and N. Mowlavi\inst{2,3} \and N.
  Produit\inst{2,3} \and S. Soldi\inst{2,3} \and R.  Walter\inst{2,3}
  \and P. Dubath\inst{2,3} \and P.  Lubi{\'n}ski\inst{4,2} \and M.
  T\"urler\inst{2,3} \and W. Coburn\inst{5} \and A.
  Santangelo\inst{1} \and R. E.  Rothschild\inst{5} \and
  R. Staubert\inst{1}}

\offprints{I.  Kreykenbohm,\\ e-mail: Ingo.Kreykenbohm@obs.unige.ch}

\institute{Institut f\"ur Astronomie und Astrophysik -- Astronomie,
  Sand 1, 72076 T\"ubingen, Germany \and INTEGRAL Science Data Centre,
  16 ch.\ d'\'Ecogia, 1290 Versoix, Switzerland \and Observatoire de
  Gen{\`e}ve, Chemin des Maillettes 51, 1290 Sauverny, Switzerland 
  \and N. Copernicus Astronomical Center, Bartycka 18, Warsaw 00-716,
  Poland \and Space Sciences Laboratory, University of California,
  Berkeley, Berkeley, CA, 94702-7450, U.S.A. \and Center for
  Astrophysics and Space Sciences, University of California at San
  Diego, La Jolla, CA 92093-0424, U.S.A.  }

\date {Received: --- / Accepted: ---}

\abstract{We present the analysis of a 100\,ksec \integral
  (3--100\,\kev) observation of the transient X-ray pulsar \V in
  outburst. The source is pulsating at
  $P_\text{Pulse}=4.3751\pm0.0002$\,s with a clear double pulse from
  6\,\kev to 60\,\kev.  The average flux was \ca550\,mCrab between
  20\,\kev and 60\,\kev.  We modeled the broad band continuum from
  5\,\kev to 100\,\kev with a power-law modified by an exponential cut
  off.  We observe three cyclotron lines: the fundamental line at
  24.9\err{0.1}{0.1}\,\kev, the first harmonic at
  50.5\err{0.1}{0.1}\,\kev as well as the second harmonic at
  71.7\err{0.7}{0.8}\,\kev, thus confirming the discovery of the
  harmonic lines by \citet{coburn05a} in \xte data.

\keywords{X-rays: stars -- stars: flare -- stars:
    pulsars: individual: V0332+53 -- stars: magnetic fields} }

\maketitle

\section{Introduction}
\label{intro}
\object{V\,0332$+$53}
(RA=03$^\text{h}$34$^\text{m}$59$^\text{s}\!$.89,
DEC=$+$53\grad10'23''$\!.$6) is a recurrent transient X-ray pulsar.
The system consists of the O8--9Ve star BQ~Cam
\citep{negueruela99a,bernacca84a} and a neutron star (NS) in an
eccentric ($e=0.31$) 34.25\,d orbit, spinning with a period of
4.375\,sec \citep{stella85a}.  Since its discovery by the Vela~5B
satellite during an outburst in 1973 \citep{terrell84a}, the system
has so far exhibited three more outbursts in 1983
\citep{tanaka83a}, 1989 \citep{makino89a}, and 2004.  The peak
luminosities during these outbursts were quite different: while the
1973 outburst reached 1.6\,Crab \citep{negueruela99a}, the flux of the
1983 outburst was about 10 times lower \citep{unger92a}, and the
1989 outburst reached a flux of 0.4\,Crab \citep{makishima90b}.  Based
on a distance of 7\,kpc \citep{negueruela99a}, which places the source
beyond the Perseus spiral arm in the outer arm of the galaxy, the
maximum X-ray luminosity reached during the 1973 outburst was
$L_\text{X} > 10^{38}$\,erg\,s$^{-1}$ -- close to the Eddington
luminosity.


For the 1989 outburst observed by \ginga \citep{makishima90b}, the
X-ray spectrum has been modeled by a power-law plus a high energy cut
off .  The analysis further revealed a deep and broad absorption
line at 28.5\,\kev which was interpreted as a cyclotron resonant
scattering feature (CRSF). The 1989 data also revealed the presence of
a quasi periodic oscillation (QPO) with a centroid frequency of
0.05\,Hz \citep{takeshima94a}. The source furthermore exhibits
aperiodic short term variability similar to Cygnus~X-1
\citep{stella85a}.

The optical companion BQ~Cam was observed to brighten from January
2002 on by \citet{goranskij04a}. Since a similar brightening had been
observed to precede the two previous X-ray outbursts, they predicted
an outburst of \V within the next one or two years.  Indeed, the
source entered an outburst in November 2004, after almost 15 years of
quiescence, recorded by the All-Sky Monitor \citep[\asm,][]{swank04a}
onboard the Rossi X-ray Timing Explorer (\xte). The X-ray flux
increased continuously and reached 1\,Crab in the 2--12\,\kev band in
2004 December \citep{remillard04a}.
Subsequent pointed observations with \xte revealed the presence of
three cyclotron lines, at $26.34\pm 0.03$\,\kev, $49.1\pm 0.2$\,\kev, and
$74\pm 2$\,\kev \citep{coburn05a}.

In this Letter, we report on the 3--100\,\kev analysis based on the
observation of \V performed on 2005 January 6--10 by the \integral
satellite.  We describe the \integral observations in
Sect.~\ref{Sect:observations}.  Imaging, spectral, and timing analysis
are presented in Sect.~\ref{Sect:analysis}, and summarized in
Sect.~\ref{Sect:discussion}.  

\section{\integral observations}
\label{Sect:observations}


Although the peak of the outburst of \V was reached in 2004 December,
a dedicated target of opportunity (TOO) observation could not be
scheduled before 2005 January 5 due to satellite orientation
constraints relative to the Sun.
100\,ksec of good data were recorded for \V between January 6 and 10 --
30\,ksec in staring mode in revolution 272 (pointing 75 on January 6th
08:47:45--18:14:00\,UT), 50\,ksec in hexagonal mode in revolution 273
(pointings 66 to 82 from January 8 at 22:18:12 to 9th at
15:45:47\,UT), and 20\,ksec in hexagonal mode in revolution 274
(pointings 4 to 9 on January 10 from 03:20:30 to 09:22:11\,UT). \V was
always in the field of view (FOV) of all three high energy
instruments.

\section{Data analysis}
\label{Sect:analysis}

We analyzed data from all instruments on board \integral. The X-ray
instrument JEM-X provides data in the 2--30\,\kev energy band. The
ISGRI layer of the imager IBIS covers the 20--800\,\kev band with an
imaging resolution of 2 arcmin and about 10\% energy resolution at
20\,\kev. The PICsIT layer provides data above 200\,\kev.  The
spectrometer SPI provides high energy resolution (1\,\kev below
100\,\kev) in the 20\,\kev to 8\,MeV energy range.  The
optical monitoring camera (OMC) provides the magnitude in the V-band
for selected optical targets.

Imaging, spectral, and timing analyses were performed using version
4.2 of the Offline Scientific Analysis (OSA) package as described in
the cookbooks available with the OSA.


\subsection{Image analysis}

The analysis of \V is facilitated by the fact that it is the only
bright X-ray source in the FOV. The mean flux during the three
observations is about 550\,mCrab in the 20--60\,\kev energy band in
\textsl{ISGRI} and 700\,mCrab in the 3--10\,\kev energy band in JEM-X,
consistent with \xte/\asm.
The flux at 20\,\kev during these observations is similar to that
measured in December 2004 during the Galactic Plane Scan
\citep{turler04a}.  We note that, at lower energies, the \xte/\asm has
recorded a steady decline of the 3--10\,\kev flux from a peak value of
\ca1\,Crab at the end of 2004 December to \ca0.6\,Crab at the time of
the \integral observation, the latter value being consistent with the
JEM-X observation.

The source was not detected by PICsIT;
this is actually not surprising, knowing that the spectra of accreting
X-ray pulsars are exponentially falling at higher energies, making
them difficult to detect above 100\,\kev \citep[see,
e.g.,][]{coburn02a}.

The OMC observed the optical companion of \V, BQ~Cam at $M_V \sim
15.4\pm0.2$ consistent with simultaneous ground based observations
\citep{masetti05a}. The observations confirm that BQ~Cam is in an
excited state.

\subsection{Spectral analysis}

\begin{figure}
\includegraphics[width=\columnwidth]{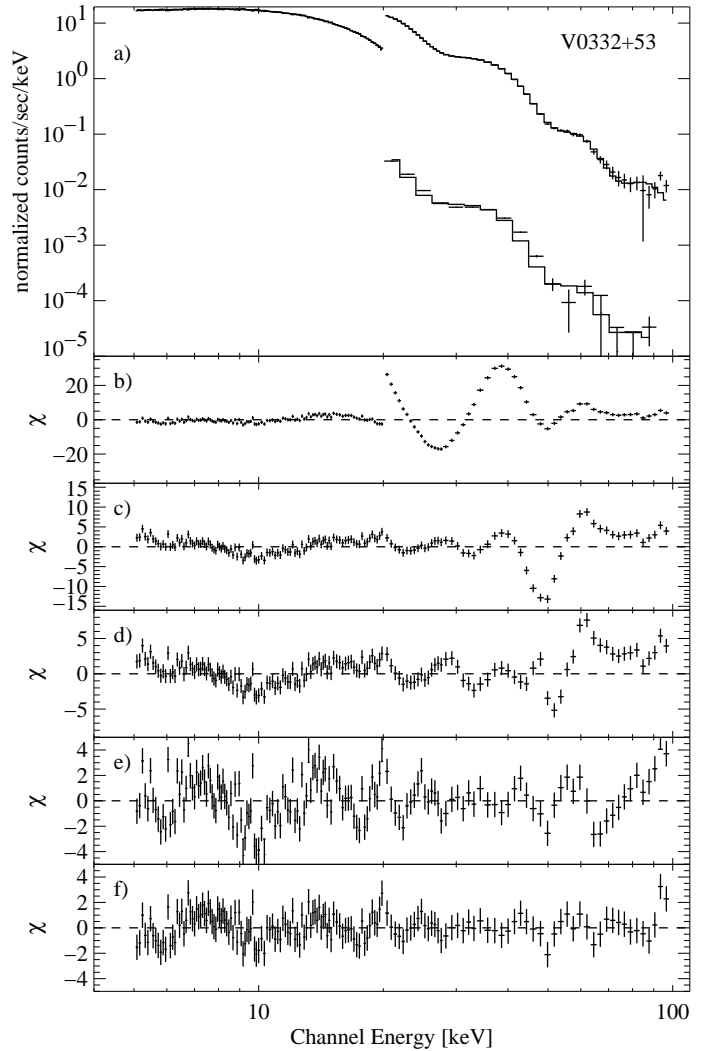}
  \caption{\textbf{a} Combined spectrum and model of data obtained
    with JEM-X (left), ISGRI (upper right), and SPI (lower right).
    Note that SPI is only shown for comparison, but is not used for
    the fits below. The dips of the cyclotron lines are already very
    apparent in the raw data.  The continuum is modeled with the
    \texttt{cutoffpl} model.  \textbf{b} model without any Gaussian
    lines applied. In \textbf{c} one Gaussian at
    26.6\err{0.1}{0.1}\,\kev is included.  In \textbf{d} another
    Gaussian is included at 47.1\err{0.1}{0.2}\,\kev.  \textbf{e} a
    third Gaussian is included 30.8\err{0.3}{0.1}\,\kev (see
    Table~\ref{fit_results}) resulting in a much better description of
    the spectrum between 20\,\kev and 40\,\kev, but significant
    deviations remain above 50\,\kev. \textbf{f} the best fit using a
    fourth Gaussian at 71.7\err{0.7}{0.8}\,\kev results in small
    residuals also at high energies. Note that all parameters
    mentioned previously are for the respective preliminary fits only.
    See Table~\ref{fit_results} for the parameters of the final fit.}
  \label{Fig:spectrum}
\end{figure}

We combined all data to obtain a spectrum with a very high signal to
noise ratio. Spectra of accreting X-ray pulsars are, however,
notoriously difficult to fit, especially if CRSFs are involved.
Special care has also to be taken that no artificial residuals are
introduced by the model \citep[as is the case for, e.g., the broken
power law and/or the high energy cut off, cf.][]{kreykenbohm99a}.  We
used the \texttt{cutoffpl} model of \textsl{XSPEC} which describes the
overall shape of the continuum reasonably well (see
Fig.~\ref{Fig:spectrum}).  We fit JEM-X and ISGRI data simultaneously
applying only a constant to allow for a free normalization between the
two instruments. A systematic error of 2\% has been applied to account
for the uncertainties of the response matrices of JEM-X and ISGRI. We
did not use SPI as the fit is completely dominated by the ISGRI
instrument above 20\,\kev anyway.  For comparison, however, we also
show the SPI data in Fig.~\ref{Fig:spectrum}a, which shows the CRSFs
at the same energies as ISGRI.

While the \texttt{cutoffpl} model fits the low energy data reasonably
well without any photoelectric absorption, large absorption line-like
structures are present above 20\,\kev. The observation of harmonically
spaced lines confirms that these features originate from cyclotron
resonant scattering.  We use a multiplicative Gaussian absorption line
at \ecyc=26.5\err{0.1}{0.1}\,\kev to fit the strong fundamental CRSF
between 20\,\kev and 30\,\kev \citep[discovered by ][]{makishima90b}.
At \ca50\,\kev the first harmonic CRSF \citep[discovered by
\xte,][]{coburn05a} is also extremely well detected (see
Fig.~\ref{Fig:spectrum}c).  After fitting both CRSFs, many residuals
still remain. This is due to the non-Gaussian shape of the
fundamental, as observed by \citet{mihara95a} in the \ginga data and
\citet{coburn05a} in \xte data.  Using Lorentzian instead of Gaussian
absorption lines does not improve the fit.  To model the fundamental,
we use a pair of Gaussians at 25.2\err{0.1}{0.1}\,\kev and
30.8\err{0.3}{0.1}\,\kev.  The significant residuals visible at
\ca70\,\kev (see Fig.~\ref{Fig:spectrum}e) are identified with the
second harmonic CRSF \citep[discovered by][]{coburn05a}. They are
fitted with a Gaussian at $E=71.7\err{0.7}{0.8}$\,\kev (see
Fig.~\ref{Fig:spectrum}f).  While the reported parameter values are
for the respective preliminary fits, the final fit parameters are
given in Table~\ref{fit_results}.

To further check the validity of our results, we used other spectral
models to fit the data including the Negative Positive EXponential
\citep[NPEX,][]{mihara95a}. In all cases, we found that three CRSFs
are present, although the actual line parameters depend on the shape
of the continuum.  We also note that the CRSFs observed at 25\,\kev
and 50\,\kev are much more significant than the deviations known to be
present in the OSA 4.2 response matrix or the energy calibration of
ISGRI between 20\,\kev and 50\,\kev; hence we are confident that the
spectral behavior is dominated by the source and not by calibration
issues.

\begin{table}
\caption{Fitted parameters from phase averaged spectra using JEM-X and
  ISGRI spectra. All errors represent
  90\% confidence.}
\label{fit_results}
\begin{tabular}{lcr@{\,}lr@{\,}l}
\hline
Parameter && \multicolumn{2}{c}{2 CRSFs} & \multicolumn{2}{c}{3 CRSFs}\\
\hline
\hline
$\Gamma$ && 
$-$0.46&\err{0.01}{0.01} &
$-$0.20&\err{0.01}{0.01} \\
\ecut & [\kevnxs] & 
6.5 & \err{0.1}{0.1} &
8.0 & \err{0.1}{0.1} \\

$E_\text{Cyc,1a}$ & [\kevnxs] & 
25.2 & \err{0.1}{0.1} &
24.9 & \err{0.1}{0.1} \\
$\sigma_\text{Cyc,1a}$ & [\kevnxs] & 
2.9 & \err{0.1}{0.1} & 
2.4& \err{0.1}{0.4} \\
$\tau_\text{Cyc,1a}$ & & 
1.25 & \err{0.01}{0.01} &
0.74& \err{0.01}{0.02} \\

$E_\text{Cyc,1b}$ & [\kevnxs] & 
30.8 & \err{0.3}{0.1} & 
29.0 & \err{0.1}{0.2} \\
$\sigma_\text{Cyc,1b}$ & [\kevnxs] & 
3.7 & \err{0.2}{0.1} &
5.0 & \err{0.1}{0.1} \\
$\tau_\text{Cyc,1b}$ & & 
0.81 & \err{0.02}{0.08} &
1.28 & \err{0.01}{0.01} \\

$E_\text{Cyc,2}$ & [\kevnxs] & 
50.1 & \err{0.1}{0.1} &
50.5 & \err{0.1}{0.1} \\
$\sigma_\text{Cyc,2}$ & [\kevnxs] & 
7.1 & \err{0.1}{0.3} &
8.1 & \err{0.1}{0.2} \\
$\tau_\text{Cyc,2}$ & & 
1.58 & \err{0.03}{0.01} & 
2.33 & \err{0.03}{0.02} \\

$E_\text{Cyc,3}$ & [\kevnxs] & 
\multicolumn{2}{c}{--}  &
71.7 & \err{0.7}{0.8} \\
$\sigma_\text{Cyc,3}$ & [\kevnxs] & 
\multicolumn{2}{c}{--}&
6.3 & \err{0.5}{0.4} \\
$\tau_\text{Cyc,3}$ & & 
\multicolumn{2}{c}{--}&
1.8 & \err{0.5}{0.4} \\
\hline
$\chi^2$ & (DOF) & 
230.2 & \,(147) & 
170.4 & \,(144) \\
\hline
\end{tabular}
\end{table}

\subsection{Timing analysis}

To determine the period of the pulsar we barycentered the event
arrival times for ISGRI and JEM-X and derived power spectra with the
\textsl{xronos} package for each of the three parts of the observation
(see Sect.~\ref{Sect:observations}).  No correction for orbital motion
has been applied since the uncertainties of the ephemeris
\citep{stella85a} are too large to extrapolate over 20\,years and the
derivation of a new ephemeris is out of the scope of this Letter.

\begin{figure}
 \includegraphics[width=\columnwidth]{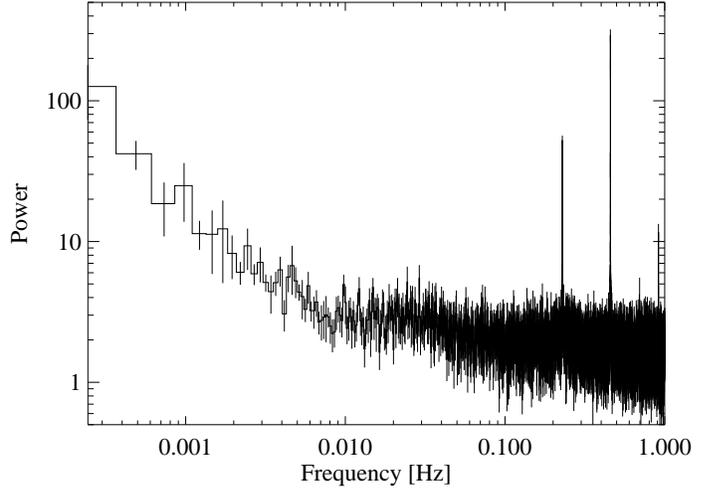}
 \caption{Power spectrum of the 50\,ksec observation in revolution 273
   of \V (see Sect.~\ref{Sect:observations}). The smaller peak at
   0.22\,Hz (4.375\,s) is due to the pulse period while the higher peak
   at 0.45\,Hz (2.188\,s) is due to the double-peaked pulse profile of
   \V (see Fig.~\ref{Fig:profile}).  }
 \label{Fig:powerspec}
 \end{figure}

Two periodicities are clearly detected in the power spectrum (see
Fig.~\ref{Fig:powerspec}): one peak at $0.2286$\,Hz associated with
the spin of the NS and one at half period due to the double-peaked
pulse profile. The derived pulse periods are $4.3753\pm0.0001$\,s
(Rev.~272), $4.37501\pm0.00006$\,s (Rev.~273), and $4.3749\pm0.0002$\,s
(Rev.~274). These variations are compatible with the effect expected
from the orbital motion.

\begin{figure}
\includegraphics[width=\columnwidth]{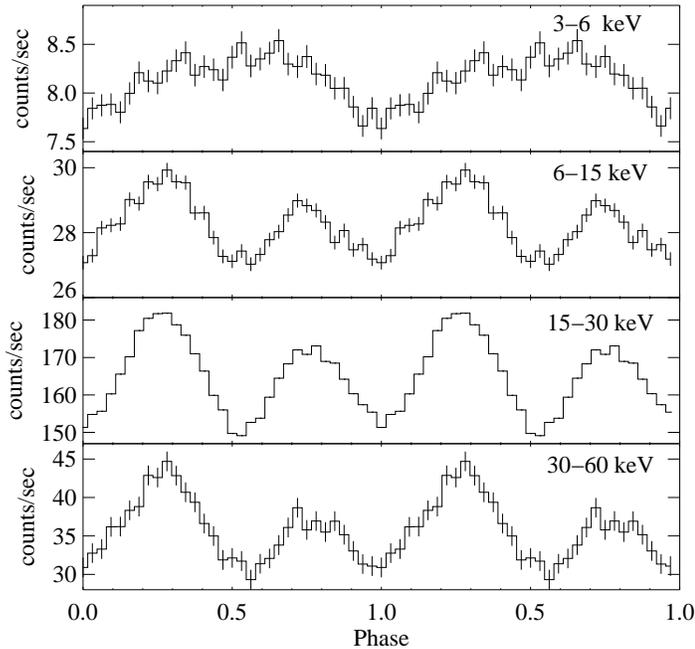}
\caption{Lightcurve of \V in two JEM-X and two ISGRI energy bands in
  Rev.~274 (see Sect.~\ref{Sect:observations}) folded with 
  $P_\text{Pulse}=4.37492$\,s. Note the change of the pulse profile at
  low energies.}
\label{Fig:profile}
\end{figure}

We created pulse profiles in several energy bands. While the pulse
profile for Rev.~272 and 273 exhibits a clear double pulse at all
energies, two behaviors are distinguishable in the pulse profile for
Rev.~274: below 6\,\kev only one pulse is detected, while two peaks are
clearly observable at all energies above 6\,\kev (see
Fig.~\ref{Fig:profile}). The secondary pulse is at a level of 60\% to
75\% amplitude of the main pulse (after subtracting the non-pulsed
flux).
At the same time, the pulsed fraction increases from 13\% in
3--15\,\kev to 22\% in 15--30\,\kev and 35\% in 30--60\,\kev.

We also performed rudimentary phase resolved spectroscopy by deriving
ISGRI spectra for the main pulse, the secondary pulse, and the pulse
minimum using data from one pointing in Rev.~274. The normalized ratio
between these spectra is consistent with unity between 20\,\kev and
45\,\kev, i.e., the spectrum in this energy range does not vary over
the pulse.  Above 45\,\kev the statistics are not sufficient from one
pointing.  More detailed pulse phase resolved spectroscopy requires a
valid ephemeris to correct for the binary motion.



\section{Summary}
\label{Sect:discussion}
\V is only the second X-ray source known so far that exhibits at least
three CRSFs; the other object, 4U\,0115$+$63, exhibits five lines
\citep{coburn04a}.  All three CRSFs in the \integral data are located
close to those energies derived from \xte.  The fundamental CRSF has
resolved structure such that even two Gaussian lines are barely able
to fit it. Comparing line shapes derived from pulse phase resolved
spectra to Monte Carlo simulations of \citet{araya99a} would allow one
to infer constraints on the geometry of the emission region.  Such a
study will be presented in a forthcoming paper.  To derive the
strength of the magnetic field, we need to derive the ``true''
centroid energy of the fundamental, which is difficult to obtain. The
first harmonic at 50\,\kev, however, is fit by a single Gaussian with
small residuals (see Fig.~\ref{Fig:spectrum}).  Assuming that the
coupling factor between the fundamental and the first harmonic is
indeed 2.0, we derive an energy for the fundamental of
$25.3\pm0.1$\,\kev, compatible with the centroid energy of the lower
of the two Gaussians used to fit the 25\,\kev feature.  If we
therefore assume that the energy of the fundamental CRSF is indeed
25.3\,\kev, that the NS has the canonical mass of 1.4\,\Msun, and that
the CRSFs originate close to surface of the NS we infer a strength of
the magnetic field of $ B = (1+z)/{11.6} \times E_\text{Cyc} \times
10^{12}\,\text{G}\approx 2.7\times 10^{12}\,\text{G}$ with a
gravitational redshift $z$ of 0.25.  The coupling factor between the
fundamental CRSF and the second harmonic is $2.85\pm0.05$, slightly
smaller than the theoretical 3.0. The same has been observed for
4U\,0115$+$63, where the shape of the fundamental line is also
strongly distorted from that of a single Gaussian and the coupling
factors are smaller than expected and is probably due to relativistic
effects \citep{heindl99b,santangelo99b}.


The derived pulse period is almost identical to the period of the 1983
outburst \cite[during which $P=4.3753\pm0.0001$,][]{stella85a},
indicating that the pulse period has, within the uncertainties arising
from the lack of orbital correction, not changed significantly since
1983.









The behavior of the source seems to be different below and above
6\,\kev. The pulse profile exhibits two peaks at all energies from
6\,\kev to 60\,\kev, similar to other accreting X-ray pulsars
\citep[e.g., \vela,][]{kreykenbohm99a}, while it exhibits only one
peak below 6\,\kev in Rev.~274 and two peaks in Rev.~272 and ~273.
The behavior is different from the 1983 outburst, where the secondary
pulse was only present at very low energies
\citep[$<5$\,\kev,][]{stella85a}.  These authors also noticed a
luminosity dependence of the pulse profile below 10\,\kev, two peaks
being detected at higher luminosities and only one at lower
luminosities. During the \integral observation, the average luminosity
in the 20--60\,\kev band was \ca550\,mCrab -- a high luminosity state
according to \citet{stella85a}, but we see a more complex behavior
with one peak below 6\,\kev and two peaks above 6\,\kev in Rev.~274.
The strength of the secondary peak is still 60\% of the main peak
between 30\,\kev and 60\,\kev.  Unfortunately, the statistics are not
sufficient to determine a pulse profile at higher energy bands.
Further analysis is in progress and will be reported in a forthcoming
paper.

\begin{acknowledgements}
  IK acknowledges DLR grants 50~OG~9601 and 50~OG~0501 and PL KBN
  grants 1P03D01827, PBZ-KBN-054/P03/2001, and 4T12E04727. RER
  acknowledges the support of NASA contracts NAS5-30720, grants
  NAG-12957 and NNG04GJ68G and of NSF international grants
  NSF\_INT-9815741 and -0003773 for fostering the UCSD/T\"{u}bingen
  scientific collaboration.
\end{acknowledgements}

\bibliographystyle{aa}
\bibliography{mnemonic,aa_abbrv,velax1,div_xpuls,xpuls,cyclotron,books,roentgen,satelliten,foreign}

\begin{thebibliography}{22}
\expandafter\ifx\csname natexlab\endcsname\relax\def\natexlab#1{#1}\fi

\bibitem[{Araya \& Harding(1999)}]{araya99a}
Araya, R.~A. \& Harding, A.~K. 1999, ApJ, 517, 334

\bibitem[{{Bernacca} {et~al.}(1984){Bernacca}, {Iijima}, \&
  {Stagni}}]{bernacca84a}
{Bernacca}, P.~L., {Iijima}, T., \& {Stagni}, R. 1984, A\&A, 132, L8

\bibitem[{Coburn {et~al.}(2002)Coburn, Heindl, Rothschild, Gruber, Kreykenbohm,
  Wilms, Kretschmar, \& Staubert}]{coburn02a}
Coburn, W., Heindl, W.~A., Rothschild, R.~E., {et~al.} 2002, ApJ, 580, 394

\bibitem[{{Coburn} {et~al.}(2004){Coburn}, {Kalemci}, {Kretschmar},
  {Kreykenbohm}, {Rothschild}, {Staubert}, \& {Wilms}}]{coburn04a}
{Coburn}, W., {Kalemci}, E., {Kretschmar}, P., {et~al.} 2004, ATel, 337, 1

\bibitem[{Coburn {et~al.}(2005)Coburn, Kretschmar, Kreykenbohm, McBride,
  Rothschild, \& Wilms}]{coburn05a}
Coburn, W., Kretschmar, P., Kreykenbohm, I., {et~al.} 2005, ATel, 381, 1

\bibitem[{{Goranskij} \& {Barsukova}(2004)}]{goranskij04a}
{Goranskij}, V. \& {Barsukova}, E. 2004, ATel, 245, 1

\bibitem[{Heindl {et~al.}(1999)Heindl, Coburn, Gruber, Pelling, Rothschild,
  Wilms, Pottschmidt, \& Staubert}]{heindl99b}
Heindl, W.~A., Coburn, W., Gruber, D.~E., {et~al.} 1999, ApJ, 521, L49

\bibitem[{Kreykenbohm {et~al.}(1999)Kreykenbohm, Kretschmar, Wilms, Staubert,
  Kendziorra, Gruber, Heindl, \& Rothschild}]{kreykenbohm99a}
Kreykenbohm, I., Kretschmar, P., Wilms, J., {et~al.} 1999, A\&A, 341, 141

\bibitem[{{Makino}(1989)}]{makino89a}
{Makino}, F. 1989, IAU Circ., 4858, 1

\bibitem[{Makishima {et~al.}(1990)Makishima, Mihara, Ishida, Ohashi, Sakao,
  Tashiro, Tsuru, Kii, Makino, Murakami, Nagase, Tanaka, Kunieda, Tawara,
  Kitamoto, Miyamoto, Yoshida, \& Turner}]{makishima90b}
Makishima, K., Mihara, T., Ishida, M., {et~al.} 1990, ApJ, 365, L59

\bibitem[{Masetti {et~al.}(2005)Masetti, Orlandini, Marinoni, \&
  Santangelo}]{masetti05a}
Masetti, N., Orlandini, M., Marinoni, S., \& Santangelo, A. 2005, ATel, 388

\bibitem[{Mihara(1995)}]{mihara95a}
Mihara, T. 1995, PhD thesis, RIKEN, Tokio

\bibitem[{{Negueruela} {et~al.}(1999){Negueruela}, {Roche}, {Fabregat}, \&
  {Coe}}]{negueruela99a}
{Negueruela}, I., {Roche}, P., {Fabregat}, J., \& {Coe}, M.~J. 1999, MNRAS,
  307, 695

\bibitem[{{Remillard}(2004)}]{remillard04a}
{Remillard}, R. 2004, ATel, 371, 1

\bibitem[{Santangelo {et~al.}(1999)Santangelo, Segreto, Giarrusso, {Dal Fiume},
  Orlandini, Parmar, Oosterbroek, Bulik, Mihara, Campana, Israel, \&
  Stella}]{santangelo99b}
Santangelo, A., Segreto, A., Giarrusso, S., {et~al.} 1999, ApJ, L85

\bibitem[{{Stella} {et~al.}(1985){Stella}, {White}, {Davelaar}, {Parmar},
  {Blissett}, \& {van der Klis}}]{stella85a}
{Stella}, L., {White}, N.~E., {Davelaar}, J., {et~al.} 1985, ApJ, 288, L45

\bibitem[{{Swank} {et~al.}(2004){Swank}, {Remillard}, \& {Smith}}]{swank04a}
{Swank}, J., {Remillard}, R., \& {Smith}, E. 2004, ATel, 349, 1

\bibitem[{{Takeshima} {et~al.}(1994){Takeshima}, {Dotani}, {Mitsuda}, \&
  {Nagase}}]{takeshima94a}
{Takeshima}, T., {Dotani}, T., {Mitsuda}, K., \& {Nagase}, F. 1994, ApJ, 436,
  871

\bibitem[{{Tanaka}(1983)}]{tanaka83a}
{Tanaka}, Y. 1983, IAU Circ., 3891, 2

\bibitem[{{Terrell} \& {Priedhorsky}(1984)}]{terrell84a}
{Terrell}, J. \& {Priedhorsky}, W.~C. 1984, ApJ, 285, L15

\bibitem[{Turler {et~al.}(2004)Turler, {Di Cocco}, Diehl, Maisala, Mas-Hesse,
  Mowlavi, Orr, Paul, \& Sunyaev}]{turler04a}
Turler, M., {Di Cocco}, G., Diehl, R., {et~al.} 2004, ATel, 372, 1

\bibitem[{{Unger} {et~al.}(1992){Unger}, {Norton}, {Coe}, \&
  {Lehto}}]{unger92a}
{Unger}, S.~J., {Norton}, A.~J., {Coe}, M.~J., \& {Lehto}, H.~J. 1992, MNRAS,
  256, 725

\end{thebibliography}

\end{document}